# LEVERAGED TRADING ON BLOCKCHAIN TECHNOLOGY

*Research in Progress*

**Johannes Rude Jensen**
University of Copenhagen
eToroX Labs
`johannesrudejensen@gmail.com`

**Victor von Wachter**
University of Copenhagen
`victor.vonwachter@di.ku.dk`

**Omri Ross**
University of Copenhagen
eToroX Labs
`omri@di.ku.dk`

## Abstract

We document an ongoing research process towards the implementation and integration of a digital artefact, executing the lifecycle of a leveraged trade with permissionless blockchain technology. By employing core functions of the 'Dai Stablecoin system' deployed on the Ethereum blockchain, we produce the equivalent exposure of a leveraged position while deterministically automating the monitoring and liquidation processes. We demonstrate the implementation and early integration of the artefact into a hardened exchange environment through a microservice utilizing standardized API calls. The early results presented in this paper were produced in collaboration with a team of stakeholders at a hosting organization, a multi-national online brokerage and cryptocurrency exchange. We utilize the design science research methodology (DSR) guiding the design, development, and evaluation of the artefact. Our findings indicate that, while it is feasible to implement the lifecycle of a leveraged trade on the blockchain, the integration of the artefact into a traditional exchange environment involves multiple compromises and drawback. Generalizing the tentative findings presented in this paper, we introduce three propositions on the implementation, integration, and implications of executing key business processes with permissionless blockchain technologies. By conducting computational design science research, we contribute to the information systems discourse on the applied utility of permissionless blockchain technologies in finance and beyond.

## Introduction

Leveraged trading is the practice of amplifying the buying power in a portfolio of assets, by trading with borrowed funds. Leveraged, or margin trading, is typically employed by traders in bridging liquidity gaps or towards the aim of achieving additional exposure to selected assets. A leveraged trade is generally executed as a non-discretionary unilateral agreement between a trader and a brokerage platform in which the trader is issued a short-term loan against the value of a collateral or margin account. The trader maintains a collateral account with the brokerage, against which the brokerage facilitates leveraged trading, either directly on the spot markets or by utilizing a set of regulated derivatives such as contracts-for-difference (CFD) or futures contracts of variable maturities. In most cases, the trader can continue trading with borrowed funds, so long as the value of the collateral or margin account exceeds any losses incurred through the depreciation of the trader's open positions.

If, at any point, the value of the open positions held by the trader drops below a certain lower bound, the brokerage or lender will typically issue a *margin call*. If, after one or more attempts, the margin call is not met, the brokerage or lender will seek to *liquidate* the trader, by closing positions in the trader's portfolio while withholding any collateral assets to recoup the initial loan. Brokerage and exchange platforms have emerged as powerful players in leveraged or margin trading, processing a significant portion of the +$60 trillion global volumes traded in equities annually.

In the highly volatile crypto-currency markets, traders with a significant appetite for risk can find exorbitant leverage multiples on unregulated exchanges, multiplying the buying power of their capital with up to excessive amounts. Because unregulated cryptocurrency exchanges can unilaterally control the execution logic of the agreement, cases have emerged in which covert service providers have offered derivative transactions with 100x leverage multiples and beyond[1], capitalizing on aggressive liquidation systems in which the trader's margin or collateral assets are unfairly seized prior to an actual liquidation event[2].

In this paper, we document an ongoing effort towards examining the feasibility of implementing the lifecycle of a leveraged position, with blockchain technology. Utilizing the design science research methodology, we design, develop, and evaluate a digital artefact comprising trading infrastructure for blockchain-based decentralized leverage. We utilize the Dai stablecoin system deployed on the Ethereum blockchain to execute the full lifecycle of a leveraged trade in the deterministic and transparent computational environment afforded by the Ethereum blockchain. The artefact is the result of an ongoing development process conducted between the authors and a team at a hosting organization, a leading international brokerage platform. We address the research question: *To what extent can blockchain technology improve the execution of a leveraged trade?* To this end we (i) demonstrate an implementation of a leveraged trade on the Ethereum blockchain using the Dai stablecoin system (ii) demonstrate and discuss the efforts towards integrating the implementation in a traditionally 'hardened' exchange infrastructure, and (iii) evaluate the challenges and compromises required in the implementation and integration of permissionless blockchain technologies in the enterprise setting.

Generalizing the findings produced in this ongoing work for future IS research, we present three propositions on the implementation, integration, and impact of permissionless blockchain technology. By conducting computational design science (Rai 2017) we aim to contribute novel insights towards the growing discourse on the design processes for digital artefacts utilizing blockchain technologies. Specifically, we aim to stimulate further discourse on the discrepancies between permissionless technologies and enterprise infrastructure. We maintain that design driven and empirical IS research is vital in generating applied perspectives on the efficacy of emerging IT infrastructure. Given the aptitude for interdisciplinary and problem-oriented scholarship, the IS community is exceptionally well positioned to explore the potential of innovative digital technologies in the financial industries and beyond.

---

[1] https://www.cftc.gov/PressRoom/PressReleases/8270-2

[2] https://www.bitmex.com/app/liquidation#Liquidation-Process

# 1    Literature Review

Blockchain technology has been of evident interest to the IS community for several years, yielding multiple novel contributions of a theoretical or design driven nature (Labazova 2019). Within the financial services, scholars have explored the implementation of blockchain technology in KYC data sharing (Moyano and Ross 2017) accounting practices (Dai and Vasarhelyi 2017) post-trade processing (Ross et al. 2019) and the execution of financial contracts (Egelund-Müller et al. 2017). Beyond the purview of the financial services, scholars have pursued innovations in supply chain management (Müller-Bloch et al. 2017) and energy markets (Castellanos et al. 2017), and beyond.

Through the pioneering efforts of early IS scholarship, the community is now in possession of multiple theoretically exhaustive taxonomies delineating the commercial implications of blockchain technologies in variety of settings (Beinke and Nguyen Ngoc 2018; Glaser 2017) juxtaposed by critical examinations of the practical capabilities of the technology (Pedersen et al. 2019). The growing body of theoretical literature in the field has led senior scholars to call for design driven research, exploring the practical implications of blockchain and smart contract technologies, in and outside the financial services and the enterprise setting (Lindman et al. 2017; Rossi et al. 2019).

A permissionless blockchain is a type of distributed database architecture, in which a singleton state machine is maintained amongst a distributed network of nodes. State changes to the shared database require consensus amongst a majority of active nodes and cannot be changed once a transaction is submitted. Because all computations must be replicated amongst a distributed set of nodes, permissionless blockchain technology can be considered *deterministic* and *transparent*. Recent iterations of the technology have introduced a virtual machine with a higher-level programming language into the deterministic execution environment. Users can write scripts commonly referred to as "smart contracts" (Antonopoulos and Wood 2018) facilitating the transparent and deterministic execution of a given business logic. The Dai stablecoin system is amongst the first and most successful smart contract-based applications on the Ethereum blockchain. Since its inception, the project has exhibited tremendous growth holding a $2.39bn valuation of assets locked at the time of writing, with the market capitalization of the associated governance token at a $548m market capitalization, down from an all-time-high surpassing $1bn.

The most recent iteration of the Dai stablecoin system, *Multi-Collateral Dai* (MCD), is a smart contract system, designed to issue the collateral-backed stablecoin, 'Dai'. The smart contract system accepts selected tokenized assets (Ross et al. 2019) and subjects the price of Dai to an algorithmic stability mechanism by which the asset exhibits a floating peg to the US dollar.[3] To withdraw Dai, a trader must collateralize a sum of tokenized assets within the smart-contract system. The collateral value guarantees the outstanding Dai, effectively 'borrowed' to the trader by the MCD smart contracts. The smart contract type responsible for accepting tokenized assets and issuing Dai is called a 'vault'. To cover the credit risk exposure accepted by the MCD contracts, a 'vault' must always be *over-collateralized* by a ratio reflecting the perceived risk and volatility of the collateral asset. The MCD contracts utilize a sophisticated set of pricing oracles to continuously monitor the market price of the assets collateralized in vaults. Throughout the lifecycle of a vault, the trader can add or remove collateral assets, ensuring that the value of the collateral assets prevails above the liquidation price, at which the 'loan' will be liquidated, and the collateral auctioned away to arbitrageurs for Dai at prices marginally below market value.

# 2    Methodology and Artefact Requirements

The iteration of the artefact presented in this paper is the result of a 6-month development process involving key stakeholders from a hosting organization. The hosting organization is a major international online brokerage with a long-term strategic interest in blockchain technology. Working with the participating stakeholders, we selected the design science research (DSR) methodology (Gregor and Hevner 2013) guiding a "build-demonstrate-evaluate" workflow in which multiple cyclical iterations of the artefact were built, tested and evaluated (Fridgen, Urbach and Schweizer, 2017). We defined a six-step process, denoting a feedback loop between development and evaluation of core components of the artefact.

---

[3] We have chosen not to elaborate on the stability mechanism stabilizing the value of Dai in this section, for resources on this topic, we recommend visiting: https://docs.makerdao.com/.

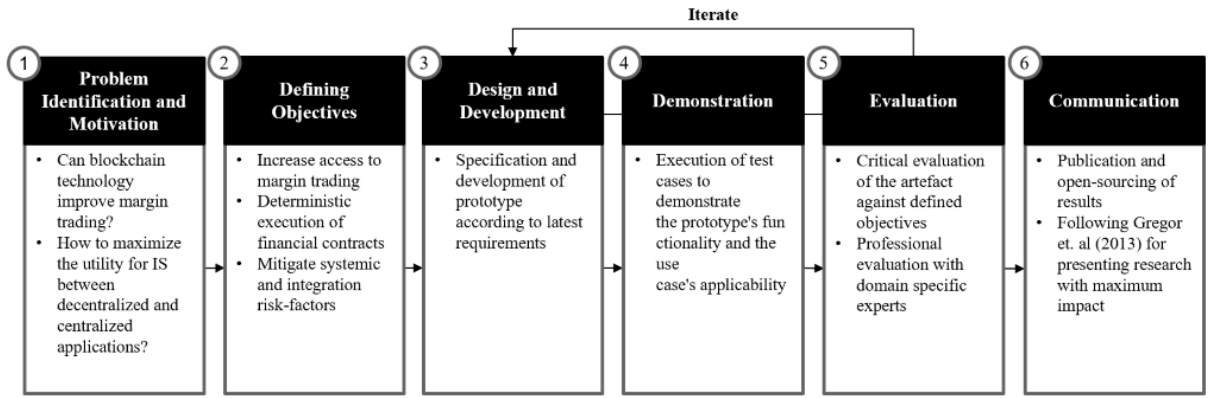

*Figure 1.*      *The design science research workflow defining the project*

The requirements for the present iteration of the artefact were gathered through multiple unstructured qualitative interviews with the stakeholders throughout the cyclical development process (Myers and Newman 2007). The authors of this paper played an integral role in the design and development of the artefact, working with the team of stakeholders in various roles (Table 1.).

| Stakeholder | Contribution |
|---|---|
| Head of Department | Strategic guidance. |
| Director of R&D | Integration and implementation guidance. |
| Managing director (author) | Managerial oversight and resource allocation. |
| Product manager (author) | Product overview, planning and execution. |
| Software development team (author) | Software developers and security architects. |
| MakerDAO software development team | Implementation assistance, support and feedback. |

*Table 1.*      *Stakeholders involved with the present iteration of the artefact*

Through multiple iterations of the cyclical workflow, we defined a set of three tentative requirements, applied to the iteration of artefact presented in this paper (Table 2).

| Requirement | Description |
|---|---|
| Deterministic execution of the leveraged trade lifecycle | The artefact must execute the lifecycle of a leveraged trade deterministically on the blockchain. |
| Non-custodial execution of the leveraged trade lifecycle | The artefact must execute the opening, maintenance and closing or liquidation of a leveraged position, without subjecting the trader to custodial risk. |
| Non-invasive integration of artefact into hardened exchange environment | The artefact must integrate with the exchange and be accessible from the exchange interface without exposing sensitive infrastructure. |

*Table 2.*      *The artefact requirements posed for the present iteration of the he artefact*

# 3  The Implementation and Integration of the Artefact

The present iteration of the artefact comprises a managed microservice running in isolation from the core platform infrastructure. The microservice runs a customized open-source library acting as a wrapper around the Dai stablecoin system smart contracts, enabling traders to interact with the smart contract systems directly through the exchange interface. Using the exchange interface, traders can open, manage, monitor, and close a leveraged position on the blockchain. We utilize the Dai stablecoin system to achieve the effect of a leveraged trade, by performing a process referred to as 'recursive re-collateralization' of a vault. For this demonstration of the artefact, we use the native asset class for Ethereum, 'ETH', with a collateral requirement of 1.5 meaning that the value of the Dai withdrawn from the vault, must not exceed 66.66% of the collateral value.

To open a leveraged position, the microservice calculates and submits a single transaction on the Ethereum blockchain, executing a sequence of three operations: (I) Open a 'vault' by collateralizing ETH and withdrawing the maximum amount Dai possible (II) exchange Dai for ETH on the Oasis exchange[4] (III) re-collateralize ETH in the same vault and withdraw Dai. This sequence is repeated recursively from step II, until the desired degree of leverage is obtained. The cumulative leverage of a given position converges to a finite value defined by the collateral requirement of the collateral asset. Because the ratio is constant, the recursive operation converges to a geometric series. We can compute the cumulative leverage ratio for an asset as:

$$L = \frac{1}{1 - \left(\frac{1}{r}\right)}$$

Let $r$ represent the collateral requirement and $L$ represent the theoretical maximum leverage ratio. As follows, the maximum theoretical leverage ratio for ETH is 3X (Figure 3.). It follows that the maximum leverage ratio approaches infinity as the collateral ratio approaches 1. However, owing to the computational costs[5] on the Ethereum network combined with the transaction fees and liquidity limitations on Oasis or UniSwap[6], successive re-collateralizations rapidly become prohibitively expensive. We estimate the practically feasible maximum leverage for a standard trade to appear around a leverage ratio of 20X for assets with a collateral requirement of 1.1 (90%)[7].

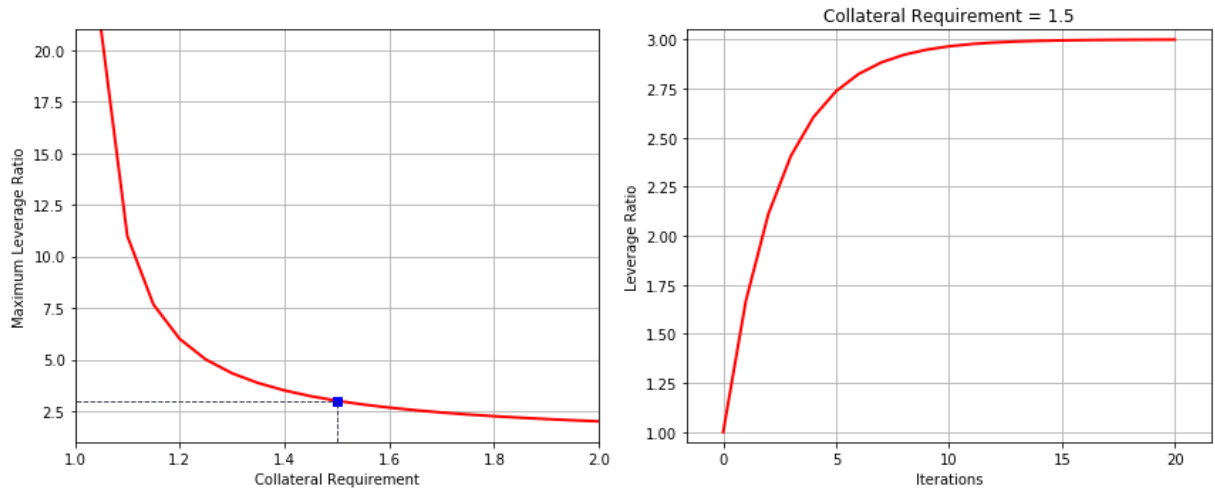

*Figure 3.  Max. leverage ratio over collateral requirements and the max. leverage ratio for ETH*

---

[4] https://oasis.app/trade/instant

[5] Referred to as "gas cost": https://etherscan.io/chart/gasprice

[6] https://app.uniswap.org/#/swap

[7] Noting that no asset class in the Dai Stablecoin system currently has a collateral ratio below 1.25 (75%)

The microservice operates under similar premises as a standard user on the exchange, utilizing the API calls to open trigger transactions from a separate, browser-based wallet, in which the user is required to maintain the initial collateral amount required to fund the leveraged position. We discern between five 'layers' of the integration (Table 3.).

| Component | Role |
|---|---|
| User interface | Visual wrapper around hardened exchange operation |
| Exchange operation | The hardened exchange environment in which the standard trading operations takes place. |
| Microservice | Communicates with the exchange operation through standardized API calls and runs four operations (i) opening a vault (ii) reading the state on the blockchain (iii) collateralizing a vault (iv) closing a vault. |
| Browser-based wallet | Is managed by the user and contains the funds required to initialize the leveraged position. |
| Ethereum blockchain | executes the deterministic automation of the monitoring and liquidation flow leveraged position by utilizing the Dai Stablecoin system to recursively re-collateralize a vault. |

*Table 3.        The five components in the integration*

The microservice runs the key operations required to open a vault with a user-defined leverage multiple by submitting a transaction with the correctly encoded input data to be signed by the user's browser-based wallet. Once the vault is open, the microservice monitors the position by reading the state of the blockchain and reporting to the exchange operation where P/L calculations are conducted against a rate database which defines the margin-call values for the account. Reading the state of the blockchain also enables the microservice to report liquidated positions to the user. The user can re-collateralize or close a position by submitting transactions through the interface, encoding a transaction through the browser-based wallet.

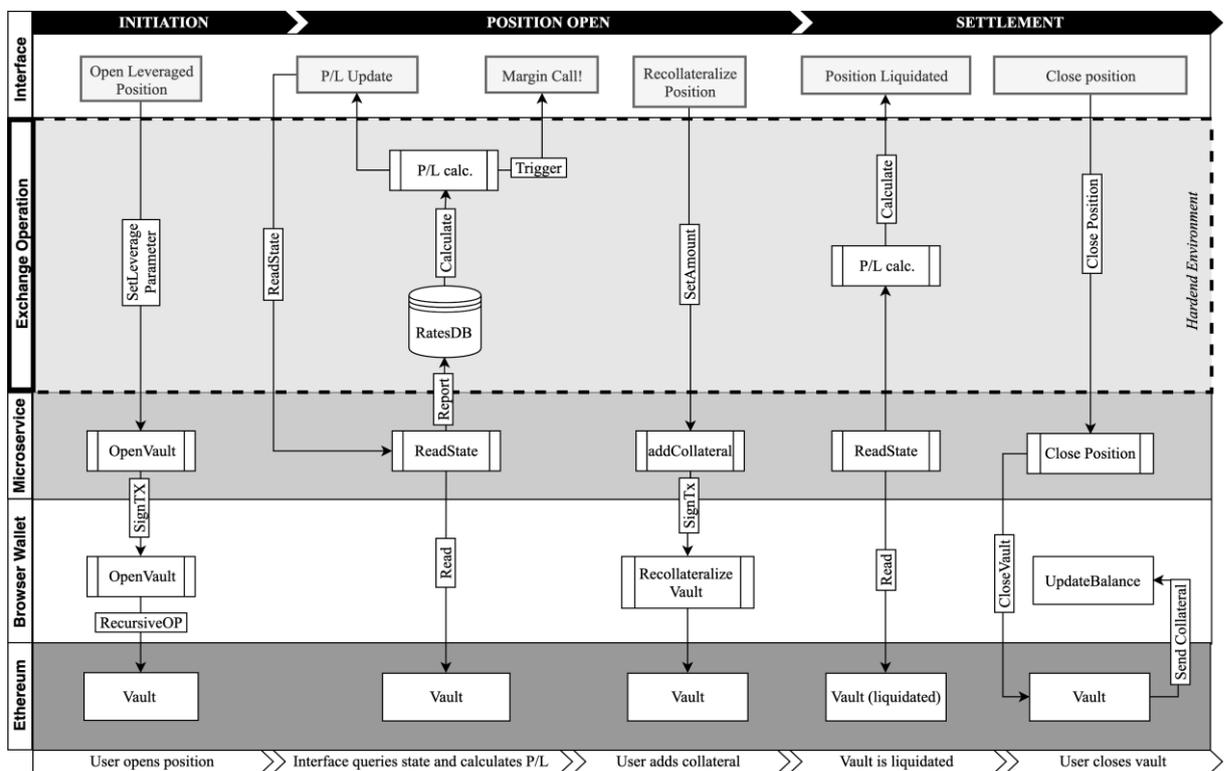

*Figure 2.        Integration taxonomy emphasizing key operations*

# 4 Artefact Evaluation

As evident in the artefact demonstration, the artefact successfully executes the lifecycle of a leveraged trade in the deterministic computational environment afforded by the Ethereum blockchain. This is achieved by interfacing with the Dai stablecoin system through a microservice acting as a wrapper around the smart contracts. We present the summarized results of the final evaluation cycle, leading to the present iteration of the artefact (Table 3.)

| Requirement | Description |
|---|---|
| Deterministic execution of the full lifecycle of a leveraged trade. | By utilizing the Dai stablecoin system to create the effect of a leveraged trade, the artefact successfully replicates the exposure of a leveraged trade entirely on the blockchain while deterministically executing the business logic required to monitor and liquidate a position. |
| Non-custodial execution of the full lifecycle of a leveraged trade. | The non-custodial integration of the microservice completes the process of recursive collateralization successfully without taking custody of user funds. All processes related to monitoring and liquidation a leveraged position are deterministically automated. |
| Non-invasive integration into exchange infrastructure | The integration was successfully completed using standard API calls combining a microservice and external browser-based wallet software. |

*Table 3. Evaluation of the artefact requirements*

While the requirements for the present iteration were tentatively met, the evaluation reveals significant drawbacks in the artefact design. First, the Dai stablecoin system and the Ethereum blockchain introduces several domain-specific limitations, as the maximum obtainable leverage ratio is constrained by the collateral requirement for the asset class, and position sizes are constrained by transaction fees and on-chain liquidity constraints in decentralized exchanges.

Second, requirements two and three introduces significant challenges for the integration of the artefact. Compliance with the requirement for a non-invasive integration using standardized API calls means that the present iteration of the artefact cannot access user funds through the hardened exchange environment, as this level of access would compromise the industry mandated cold-storage requirements. Instead, the implementation requires that the user holds sufficient funds in an associated browser-based wallet which can expose an API[8] enabling the signature of the transaction input data comprising the recursive logic which creates the leveraged position. To fund such a wallet, the user must first download and install the wallet software in her browser and subsequently connect the wallet with the exchange and fund the wallet by transferring the correct amount of assets from the exchange (or elsewhere) to the wallet. Additionally, the use of an external browser-based wallet implies that the user must carry the transaction costs for the recursive operation. As the recursive operation utilizes a decentralized exchange (DEX) instead of the exchange orderbook itself, liquidity provisioning for the artefact is externalized.

# 5 Discussion

While the present iteration of the artefact successfully executes the lifecycle of a leveraged trade on the blockchain, the evaluation exposed several weaknesses in the artefact design. These issues were primarily motived by the dissimilarities between a hardened environment such as an exchange database system and a permissionless computational environment, such as the Ethereum blockchain. The tentative achievements presented here came at the cost of several compromises all of which ultimately affects the potential competitiveness of a product of this, or similar, nature. Generalizing the findings produced in this work for a broader IS audience, we offer three propositions on the integration, implementation, and impact of permissionless blockchain technology in the enterprise setting.

---

[8] We used MetaMask for the implementation: https://metamask.io/

**Proposition 1:** *The integration of permissionless blockchain technology in the enterprise setting introduces novel practical and theoretical contradictions.*

While the execution of key business processes with permissionless blockchain technology has been shown to introduce performance enhancements and improved risk mitigation across several use-cases in the financial industries (Labazova 2019) the integration of permissionless infrastructure in hardened enterprise infrastructure is likely to introduce ambiguities between traditional and standardized approach to risk management in siloed database architectures, and the radical 'openness' of blockchain technology (Schlagwein et al. 2017).

**Proposition 2:** *Client facing implementations of digital artefacts utilizing permissionless blockchain technologies succeed by assimilating relevant parts of the value chain to the key properties of permissionless blockchain technology.*

The successful integration of digital artefacts implemented with permissionless blockchain technologies, ought to assimilate necessary aspects of the value chain to the key properties of blockchain technology: Transparency, deterministic execution, and non-custodial management of tokenized assets. Attempts at assimilating business processes executed with blockchain technology into a legacy environment may translate into lack of competitiveness and redundancy. Our advice is to build from the blockchain and up, not the other way around.

**Proposition 3:** *The dissemination of permissionless blockchain technology and the deterministic execution of critical business processes introduces a level playing field for applications, users, and service providers.*

An interesting property of permissionless blockchain technology as a computational environment, is the equivalency of computations. Unlike cloud computing models where organizations operate in secured walled environments (Weinhardt et al. 2009), permissionless blockchains introduce a secondary market for transaction processing denominated in the native asset-class.[9] Consequently, computations are prioritized by transaction fees which are, in most cases, paid by the user. The shared computational environment for blockchain based applications is likely to introduce novel implications for competing organizations, as service providers utilizing the same blockchain must share the same resource-constrained computational environment.

# 6    Conclusion and Future Work

In this paper, we present ongoing efforts towards the research question: *To what extent can blockchain technology improve the execution of a leveraged trade?* Utilizing the DSR methodology we demonstrate the implementation and initial integration of a digital artefact, utilizing the Dai Stablecoin system to deterministically automate the processes required to monitor and liquidate a leveraged position. The initial attempt at integrating the artefact into a traditional exchange operation presented several necessary compromises required to maintain the integrity of the secured exchange environment.

Future work on this artefact will seek to bridge the chasm between permissionless blockchain technology and the hardened enterprise environment. Our work will examine the feasibility of exposing the browser-based wallet API to internally managed hot-wallets alongside an examination of whether the recursive operation can be conducted utilizing internal liquidity.

---

[9] The native asset for computations on the Ethereum blockchain is called 'gas' and is denominated in a floating exchange with ETH.

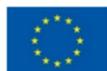 This project has received funding from the European Union's Horizon 2020 research and innovation programme under the Marie Skłodowska-Curie grant agreement No 801199

## References

Antonopoulos, A. M., and Wood, G. 2018. *Mastering Ethereum: Building Smart Contracts and Dapps*, O'Reilly Media.

Beinke, J. H., and Nguyen Ngoc, D. 2018. "Towards a Business Model Taxonomy of Startups in the Finance Sector Using Blockchain," *39th International Conference on Information Systems* (December), pp. 1–9.

Castellanos, J. A. F., Coll-Mayor, D., and Notholt, J. A. 2017. "Cryptocurrency as Guarantees of Origin: Simulating a Green Certificate Market with the Ethereum Blockchain," in *2017 IEEE International Conference on Smart Energy Grid Engineering (SEGE)*, IEEE, pp. 367–372.

Dai, J., and Vasarhelyi, M. A. 2017. "Toward Blockchain-Based Accounting and Assurance," *Journal of Information Systems* (31:3), pp. 5–21.

Egelund-Müller, B., Elsman, M., Henglein, F., and Ross, O. 2017. "Automated Execution of Financial Contracts on Blockchains," *Business and Information Systems Engineering* (59:6), pp. 457–467.

Fridgen, G., Urbach, N., and Schweizer, A. 2017. "Unchaining Social Businesses-Blockchain as the Basic Technology of a Crowdlending Platform IT-Management Im Zeitalter Der Digitalisierung View Project."

Glaser, F. 2017. "Pervasive Decentralisation of Digital Infrastructures: A Framework for Blockchain Enabled System and Use Case Analysis," *Proceedings of the 50th Hawaii International Conference on System Sciences (HICSS-50)*, pp. 1543–1552.

Gregor, S., and Hevner, A. R. 2013. "Positioning and Presenting Design Science Research for Maximum Impact," *MIS Quarterly* (37:2), pp. 337–356.

Ken Peffers, Tuure Tuunanen, Marcus A. Rothenberger, and Samir Chatterjee. 2007. "A Design Science Research Methodology for Information Systems Research," *Journal of Management Information Systems* (24:3), pp. 45–77.

Labazova, O. 2019. "Towards a Framework for Evaluation of Blockchain Implementations," in *Fortieth International Conference on Information Systems*.

Lindman, J., Tuunanen, V. K., and Rossi, M. 2017. "Opportunities and Risks of Blockchain Technologies: A Research Agenda," in *Proceedings of the 50th Hawaii International Conference on System Sciences (2017)*, pp. 1533–1542.

Moyano, J. P., and Ross, O. 2017. "KYC Optimization Using Distributed Ledger Technology," *Business & Information Systems Engineering* (59:6), Springer, pp. 411–423.

Müller-Bloch, C., Beck, R., and Palmund, S. 2017. "Blockchain to Rule the Waves-Nascent Design Principles for Reducing Risk and Uncertainty in Decentralized Environments," *Proceedings of the 38th International Conference on Information Systems* (September), pp. 1–16.

Myers, M. D., and Newman, M. 2007. "The Qualitative Interview in IS Research: Examining the Craft," *Information and Organization* (17:1), pp. 2–26.

Pedersen, A. B., Risius, M., and Beck, R. 2019. "Blockchain Decision Path: When to Use Blockchains? Which Blockchains Do You Mean?," *MIS Quarterly Executive* (18:2), p. 24. (https://pure.itu.dk/ws/files/83594249/MISQe_BC_in_the_Maritime_Shipping_Industry_Revision.pdf).

Rai, A. 2017. "Diversity of Design Science Research," *MIS Quarterly* (41:1), iii–xviii.

Ross, O., Jensen, J., and Asheim, T. 2019. "Assets under Tokenization: Can Blockchain Technology Improve Post-Trade Processing?," in *Fortieth International Conference on Information Systems,*

*Munich 2019.*

Rossi, M., Mueller-Bloch, C., Thatcher, J. B., and Beck, R. 2019. "Blockchain Research in Information Systems: Current Trends and an Inclusive Future Research Agenda," *Journal of the Association for Information Systems* (20:9), pp. 1388–1403.

Schlagwein, D., Conboy, K., Feller, J., Leimeister, J. M., and Morgan, L. 2017. "'openness' with and without Information Technology: A Framework and a Brief History," *Journal of Information Technology* (32:4), Palgrave Macmillan UK, pp. 297–305.

Weinhardt, C., Anandasivam, A., Blau, B., Borissov, N., Meinl, T., Michalk, W., and Stößer, J. 2009. "Cloud Computing – A Classification, Business Models, and Research Directions," *Business & Information Systems Engineering* (1:5), pp. 391–399.